\documentclass[aps,pra,twocolumn,groupedaddress,floatfix]{revtex4}%
\usepackage{graphicx}
\usepackage{dcolumn}
\usepackage{bm}
\usepackage{hyperref}
\usepackage{amssymb}
\usepackage{amsmath}
\usepackage{amsfonts}
\usepackage{amssymb}
\usepackage{color}%
\begin{document}
\title{Observation of magnetically tunable Feshbach  resonances in ultracold $^{23}$Na$^{40}$K+$^{40}$K collisions}
\author{Huan Yang}
\thanks{These authors contributed equally to this work.}
\affiliation{Shanghai Branch, National Laboratory for Physical Sciences at Microscale and
Department of Modern Physics, University of Science and Technology of China,
Hefei, Anhui 230026, China}
\affiliation{CAS Center for Excellence and Synergetic Innovation Center in Quantum
Information and Quantum Physics, University of Science and Technology of
China, Shanghai 201315, China}
\affiliation{CAS-Alibaba Quantum Computing Laboratory, Shanghai 201315, China}
\author{De-Chao Zhang}
\thanks{These authors contributed equally to this work.}
\affiliation{Shanghai Branch, National Laboratory for Physical Sciences at Microscale and
Department of Modern Physics, University of Science and Technology of China,
Hefei, Anhui 230026, China}
\affiliation{CAS Center for Excellence and Synergetic Innovation Center in Quantum
Information and Quantum Physics, University of Science and Technology of
China, Shanghai 201315, China}
\affiliation{CAS-Alibaba Quantum Computing Laboratory, Shanghai 201315, China}
\author{Lan Liu}
\thanks{These authors contributed equally to this work.}
\affiliation{Shanghai Branch, National Laboratory for Physical Sciences at Microscale and
Department of Modern Physics, University of Science and Technology of China,
Hefei, Anhui 230026, China}
\affiliation{CAS Center for Excellence and Synergetic Innovation Center in Quantum
Information and Quantum Physics, University of Science and Technology of
China, Shanghai 201315, China}
\affiliation{CAS-Alibaba Quantum Computing Laboratory, Shanghai 201315, China}
\author{Ya-Xiong Liu}
\affiliation{Shanghai Branch, National Laboratory for Physical Sciences at Microscale and
Department of Modern Physics, University of Science and Technology of China,
Hefei, Anhui 230026, China}
\affiliation{CAS Center for Excellence and Synergetic Innovation Center in Quantum
Information and Quantum Physics, University of Science and Technology of
China, Shanghai 201315, China}
\affiliation{CAS-Alibaba Quantum Computing Laboratory, Shanghai 201315, China}
\author{Jue Nan}
\affiliation{Shanghai Branch, National Laboratory for Physical Sciences at Microscale and
Department of Modern Physics, University of Science and Technology of China,
Hefei, Anhui 230026, China}
\affiliation{CAS Center for Excellence and Synergetic Innovation Center in Quantum
Information and Quantum Physics, University of Science and Technology of
China, Shanghai 201315, China}
\affiliation{CAS-Alibaba Quantum Computing Laboratory, Shanghai 201315, China}
\author{Bo Zhao}
\affiliation{Shanghai Branch, National Laboratory for Physical Sciences at Microscale and
Department of Modern Physics, University of Science and Technology of China,
Hefei, Anhui 230026, China}
\affiliation{CAS Center for Excellence and Synergetic Innovation Center in Quantum
Information and Quantum Physics, University of Science and Technology of
China, Shanghai 201315, China}
\affiliation{CAS-Alibaba Quantum Computing Laboratory, Shanghai 201315, China}
\author{Jian-Wei Pan}
\affiliation{Shanghai Branch, National Laboratory for Physical Sciences at Microscale and
Department of Modern Physics, University of Science and Technology of China,
Hefei, Anhui 230026, China}
\affiliation{CAS Center for Excellence and Synergetic Innovation Center in Quantum
Information and Quantum Physics, University of Science and Technology of
China, Shanghai 201315, China}
\affiliation{CAS-Alibaba Quantum Computing Laboratory, Shanghai 201315, China}

\maketitle

\textbf{Resonances in ultracold collisions involving heavy molecules are difficult to understand, and have proven  challenging to detect. Here we report the observation of magnetically tunable Feshbach resonances in ultracold collisions between $^{23}$Na$^{40}$K molecules in the rovibrational
ground state and  $^{40}$K atoms. We prepare the atoms and molecules in various hyperfine levels of their ground states and observe the loss of molecules as a function of the magnetic field. The atom-molecule Feshbach resonances are identified by observing an enhancement of the loss rate coefficients. We have observed three resonances at approximately 101 G in various atom-molecule scattering channels, with the widths being a few hundred milliGauss.  The observed atom-molecule Feshbach resonances at ultralow temperatures probe the three-body potential energy surface with an unprecedented resolution. Our work will help to improve the understanding of complicated ultracold collisions, and open up the possibility of creating ultracold triatomic molecules.}

\vspace{0.5cm}

Understanding collisions involving molecules at the quantum level has been a long-standing goal in chemical physics \cite{Herschbach2009}. Scattering resonance is one of the most remarkable quantum phenomena and plays a critically important role in the study of collisions. It is very sensitive to both the long-range and short-range parts of the molecule interaction potential, and thus offers a unique probe of the potential energy surface (PES) governing the collision dynamics. Although scattering resonances are well known and have been the main features in ultracold atomic gases and nuclear collisions \cite{chin2010}, they have proven challenging to observe in molecule systems. Recently, significant progress has been achieved in experimentally studying resonances in cold molecular collisions involving the light particles, e.g., H$_2$, HD molecule or He atom, by means of molecular beam techniques. In the crossed-beam or merged-beam experiments, shape resonances or Feshbach resonances have been observed in atom-molecule chemical reactions \cite{Skodje2000,Qiu2006,Henson2012,Wang2013,Kim2015,Yang2015}, atom-molecule inelastic collisions \cite{Vogels2015,Bergeat2015,Klein2017}, and molecule-molecule inelastic collisions \cite{Chefdeville2013,Perreault2017}. However, in these experiments, the collision energies are still high (at Kelvin or sub Kelvin), and thus a few partial waves contribute to the scattering cross sections.

Ultracold molecules offer great opportunities to study molecular collisions in the quantum regime. At ultralow temperatures, the de Broglie wavelength of the collision partners is much larger than the range of molecular interaction potential, and only the lowest possible partial wave dominates the collision process \cite{Carr2009,Quemener2012}. Consequently, the collisions at ultracold temperatures are highly quantum mechanical. Due to the anisotropy of the PES, the collisions involving ultracold molecules may support many resonances that are contributed by the rotational and vibrational excited states \cite{Bohn2002,Mayle2012}. Therefore, it is expected that scattering resonances can be routinely observed in ultracold molecular systems. For the ultracold collisions involving light molecules, the low density of resonant states allows calculations of the scattering resonances, and a lot of Feshbach resonances in atom-molecule collisions \cite{Tscherbul2006,Hummon2011,Frye2016} and molecule-molecule collisions \cite{Bohn2002,Tscherbul2009} have been predicted. However, due to the experimental difficulties in preparing the ultracold colliding particles, these predictions have not been tested.

The situation is much more complicated for the ultracold collisions involving heavy molecules, such as the alkali-metal-diatomic molecules in the rovibrational ground state created from ultracold atomic gases \cite{Ni2008,Takekoshi2014,Molony2014,Park2015,Guo2016,Rvachov2017,seesselberg2018}. The scattering resonances involving these heavy molecules are difficult to understand, and are highly challenging to observe. For the reactive collisions, the reactions are universal and the short-range losses with a near-unity probability suppress any possible resonances \cite{Ospelkaus2010a,Quemener2012}.
For the nonreactive atom-molecule collisions, the PES is so deep that thousands of rovibrational states may contribute to the resonances. As a consequence, the density of resonant states near the threshold of the collision channel is quite high and it is not clear whether the individual resonance is resolvable \cite{Mayle2012}. In this case, the theoretical calculation of the Feshbach resonances is extremely difficult, especially when nuclear spins and external fields are considered \cite{Croft2017}.  Instead, a statistical model is adopted to explore such highly resonant scattering \cite{Mayle2012}. It predicts that at a temperature of below 1 $\mu$K, for the atom-diatomic-molecule collisions, many \emph{s}-wave Feshbach resonances with an average spacing of less than 1 Gauss should be observable. However, the experimental observation of these resonances remains elusive.

Here we report on the observation of magnetic Feshbach resonances between ultracold $^{23}$Na$^{40}$K ground-state molecules ($^{1}\Sigma$) and $^{40}$K atoms. The binding energies of $^{23}$Na$^{40}$K $(v=0)$ and $^{40}$K$_{2} (v=0)$ are about 5212 cm$^{-1}$ \cite{Park2015} and 4405 cm$^{-1}$ \cite{Ospelkaus2010a}, respectively. Therefore, $^{23}$Na$^{40}$K $(v=0,N=0)$ + $^{40}$K collision is nonreactive, because the reaction $ ^{23}$Na$^{40}$K$(v=0)$ + $^{40}$K$\rightarrow$ $^{40}$K$_2$ $(v=0)$ + $^{23}$Na is highly endothermic and is forbidden at ultracold temperatures. The atomization energy of NaK$_2$ is estimated to be 7125 cm$^{-1}$ \cite{Ziuchowski2010}, which gives rise to a deep PES. The closed-channels asymptotically to the $^{23}$Na$^{40}$K  rovibrational excited states support a lot of triatomic bound states, as illustrated in Fig. 1. Besides, the reactive closed-channels asymptotically to the $^{40}$K$_2$ rovibrational states are locally open at short range and also support many bound states. These bound states may lead to a high density of resonant states near the threshold. We prepare $^{23}$Na$^{40}$K  molecules and $^{40}$K atoms in various hyperfine levels of their ground states, and search for the resonances by measuring the loss rate of the molecules due to atom-molecule inelastic collisions as a function of the magnetic field. The appearance of Feshbach resonance is identified by observing a resonantly enhanced loss rate. We have observed three resonances at the magnetic fields of about 101 G.

\begin{figure}[ptb]
\centering
\includegraphics[width=8cm]{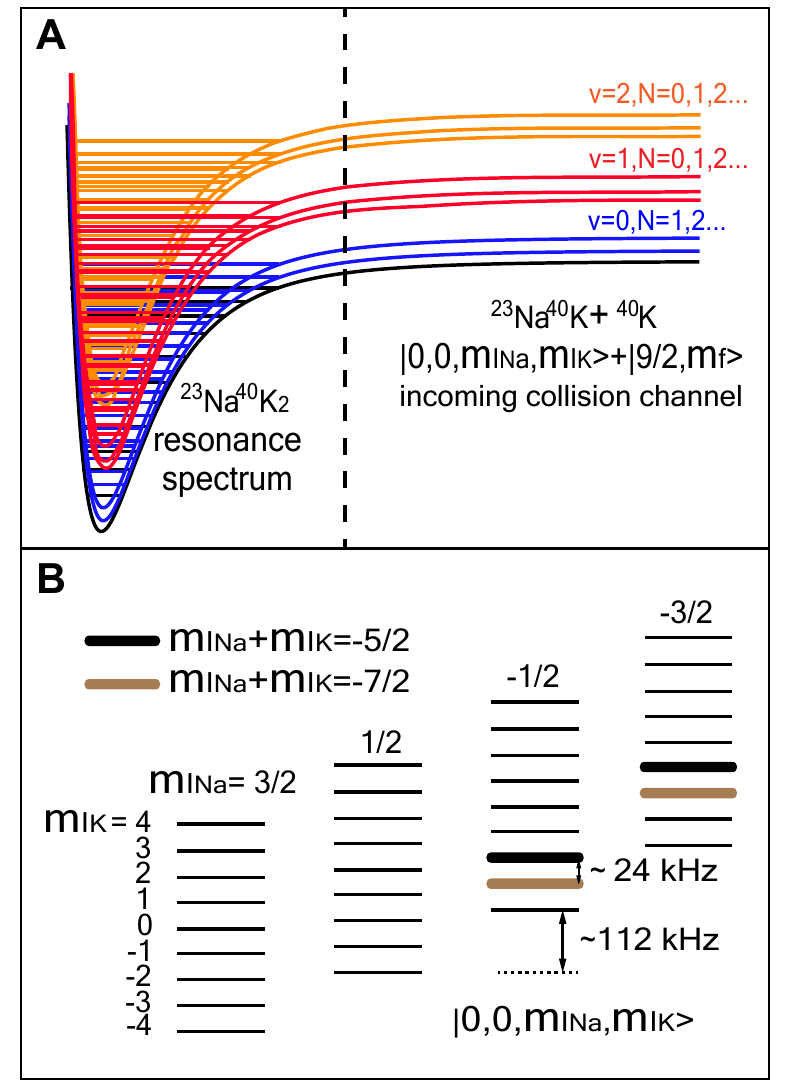}
\caption{\textbf{Illustration of the atom-molecule Feshbach resonances between ground-state $^{23}$Na$^{40}$K molecule and $^{40}$K atom}. (\textbf{A}) The potential energy surface is very deep and thus a large number of closed-channels
asymptotically to the $^{23}$Na$^{40}$K  vibrational and rotational excited states can support the triatomic bound states, which give rise to a high density of resonant states near the threshold. The incoming channel is $^{23}$Na$^{40}$K$(v=0,N=0)$+$^{40}$K in a specific combination of hyperfine states. The atom-molecule Feshbach resonances probe the short-range resonance spectrum. The energy of the collision channels can be magnetically tuned. A Feshbach resonance occurs once the energy of the incoming channel coincides with the energy of a bound state. (\textbf{B}) Hyperfine structure of $^{23}$Na$^{40}$K ground-state molecule at a magnetic field of 100 G.
The hyperfine levels of the $^{23}$Na$^{40}$K molecule in the rovibrational ground state of the $^{1}\Sigma$ singlet potential are split due to the nuclear Zeeman effects. The nuclear spin projections $m_{I_{\rm{Na}}}$ and $m_{I_{\rm{K}}}$ are approximately good quantum numbers. The hyperfine levels that are used in the current experiment are marked by thick lines. } %
\label{fig1}%
\end{figure}

Our experiment starts with ultracold $^{23}$Na and $^{40}$K atomic mixture confined in a crossed-beam optical dipole trap at a temperature of about 500 nK. There are several broad atomic Feshbach resonances between $^{23}$Na and $^{40}$K, which can in principle be employed to create the ground-state molecules and to search for the atom-molecule Feshbach resonances. There is no prior knowledge about positions or widths of atom-molecule Feshbach resonances. Only a statistical model gives a qualitative estimate the density of resonant states \cite{Mayle2012}. Although these predictions can only be considered as a qualitative guide, it indicates that there may be many atom-molecule \emph{s}-wave resonances randomly distributed with an average spacing of less than 1 G. Therefore, in a magnetic field range of a few Gauss, it may be possible to observe the Feshbach resonances.

The magnetic field where we search for the atom-molecule Feshbach resonances is in the range  99.3 $<B <103.8$ G. This magnetic field range is close to a broad atomic Feshbach resonance at 110 G. We first create weakly bound Feshbach molecules in the atom mixture by Raman photoassociation. The remaining $^{23}$Na atoms are removed immediately after the Feshbach molecules have been formed. We then transfer the molecules from the Feshbach state to the rovibrational ground state by means of the stimulated Raman adiabatic passage (STIRAP).  The details of the association and the STIRAP are given in the supplementary materials. The hyperfine levels of the ground states of the $^{23}$Na$^{40}$K molecule are labelled by $|v,N,m_{I_{\rm{Na}}},m_{I_{\rm{K}}}\rangle$, where the vibrational and rotational quantum numbers are $v=N=0$, and $m_{I_{\rm{Na}}}$ and $m_{I_{\rm{K}}}$ are the nuclear spin projections of $^{23}$Na and $^{40}$K, respectively. In our experiment, the hyperfine states $|0,0,-3/2,-2\rangle$, $|0,0,-3/2,-1\rangle$, $|0,0,-1/2,-3\rangle$, and $|0,0,-1/2,-2\rangle$ can be populated by choosing proper intermediate states and polarizations of lasers. The hyperfine structure of the ground-state molecule is shown in Fig. 1. After the ground-state molecules are prepared, the $^{40}$K atoms are transferred to different hyperfine states $|f,m_f\rangle_{\rm{K}}=|9/2,m_f\rangle$ with $m_f=-9/2,...,-1/2$ by the radio frequency  pulses. In this way, twenty different combinations of the atom and molecule hyperfine states can be prepared.

\begin{figure}[pth]
\centering
\includegraphics[width=8cm]{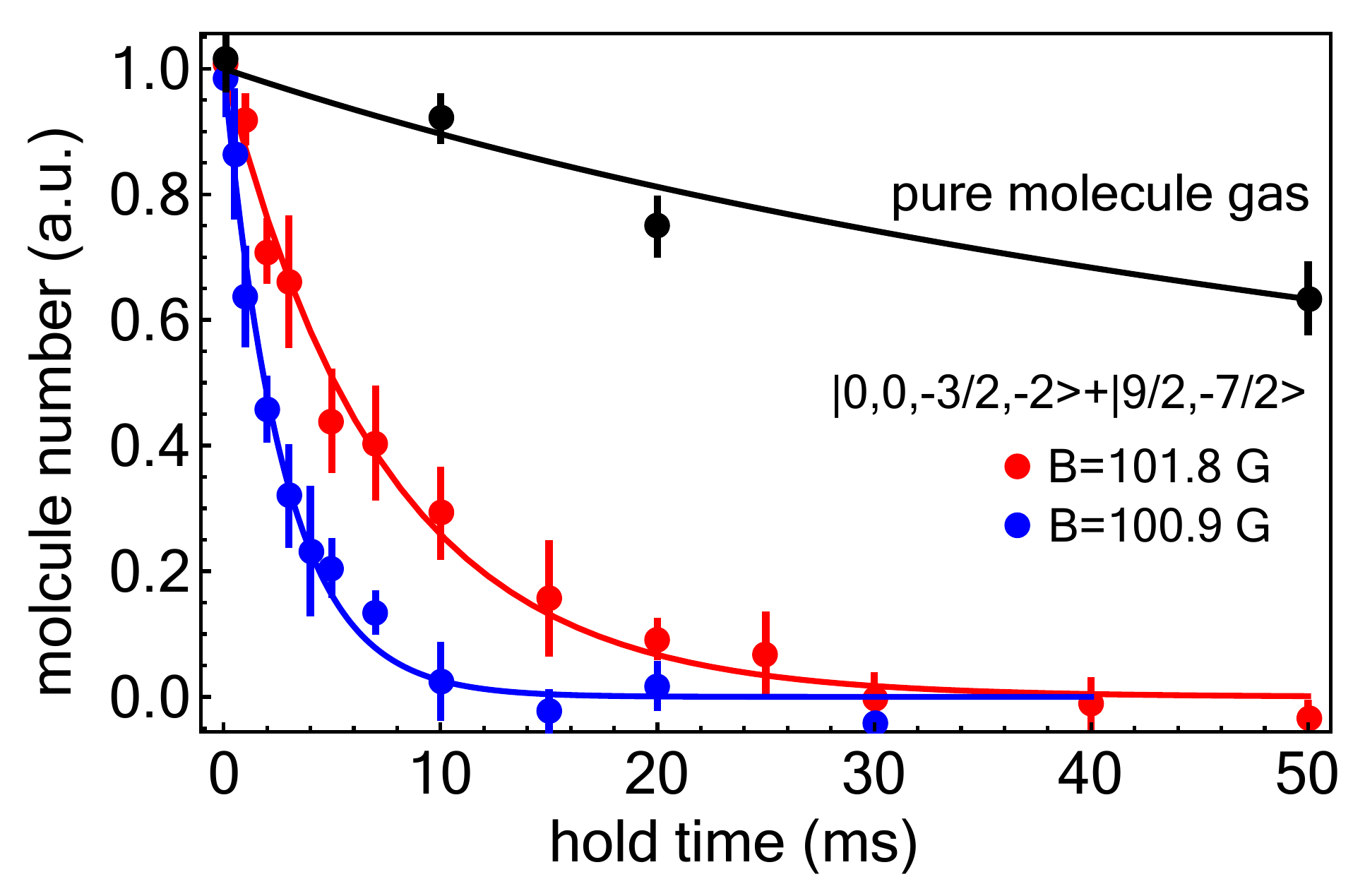}
\caption{\textbf{The decay of $^{23}$Na$^{40}$K molecule in the atom-molecule mixture}. The  time evolutions of the number of the molecules are recorded. The loss rate coefficients are extracted from the measured decay rate. As a reference, the decay of the pure molecule gas is also shown. For the $|0,0,-3/2,-2\rangle+|9/2,-7/2\rangle$ collision, it can be clearly seen that the loss rates are dependent on the magnetic field.  Error bars represent $\pm1$ s.d.}%
\label{fig2}%
\end{figure}

The $^{23}$Na$^{40}$K molecules will decay due to the two-body hyperfine-changing inelastic collisions with the $^{40}$K atoms, since the atoms and molecules are in the excited hyperfine states. We record the time evolution of the number of the molecules, as shown in Fig. 2. After a certain hold time, the number of the remaining ground-state molecules is measured by transferring  the molecules back to the Feshbach states, which are detected by the absorbtion imaging. The typical lifetime of the molecules in the atom-molecule mixture is on the order of 10 ms. This is much shorter than the lifetime of the pure molecule gas, which is larger than 100 ms. Therefore, the decay of the molecule in the mixture is dominantly caused by the $^{23}$Na$^{40}$K+$^{40}$K two-body inelastic collisions.

The decay of the molecules may be described by $\dot{N}_m=-\gamma N_m$, where $ N_m$ is the number of molecules, and $\gamma=\beta \bar{n}_a$ is the decay rate, with $\beta$ and $\bar{n}_a$ being the loss rate coefficient and the mean density of the $^{40}$K atoms, respectively. The mean atomic density may be calculated by $\bar{n}_a=((m_K \bar{\omega}^2))/(4\pi k_B T_{\rm{K}}))^{3/2}N_{a}$, with $\bar{\omega}$ the geometric mean trapping frequencies of the $^{40}$K atoms, $k_{\rm{B}}$ the Boltzmann constant, $T_{\rm{K}}$ the temperature, and $N_a$ the number of $^{40}$K atoms. In our experiment, the number of $^{40}$K atoms is about one order of magnitude larger than that of the molecules, and thus the mean density $\bar{n}_a$ is approximately a constant. In this case, the loss rate coefficient $\beta$ may be extracted from the measured decay rate $\gamma$ and the atomic mean density $\bar{n}_a$.

We search for the atom-molecule Feshbach resonances in twenty different incoming collision channels. For each channel, we measure the loss rate coefficient as a function of the magnetic field. By varying the magnetic field, we expect the energy differences between the triatomic bound states and the threshold of the incoming scattering channel will be changed. If a triatomic bound state intersects the threshold of the scattering channel and the coupling between the bound state and the scattering state is strong, a Feshbach resonance may occur. The Feshbach resonances are identified through the strongly enhanced loss rate coefficients  \cite{chin2010,Mayle2012}.

\begin{figure}[tbh]
\centering
\includegraphics[width=8cm]{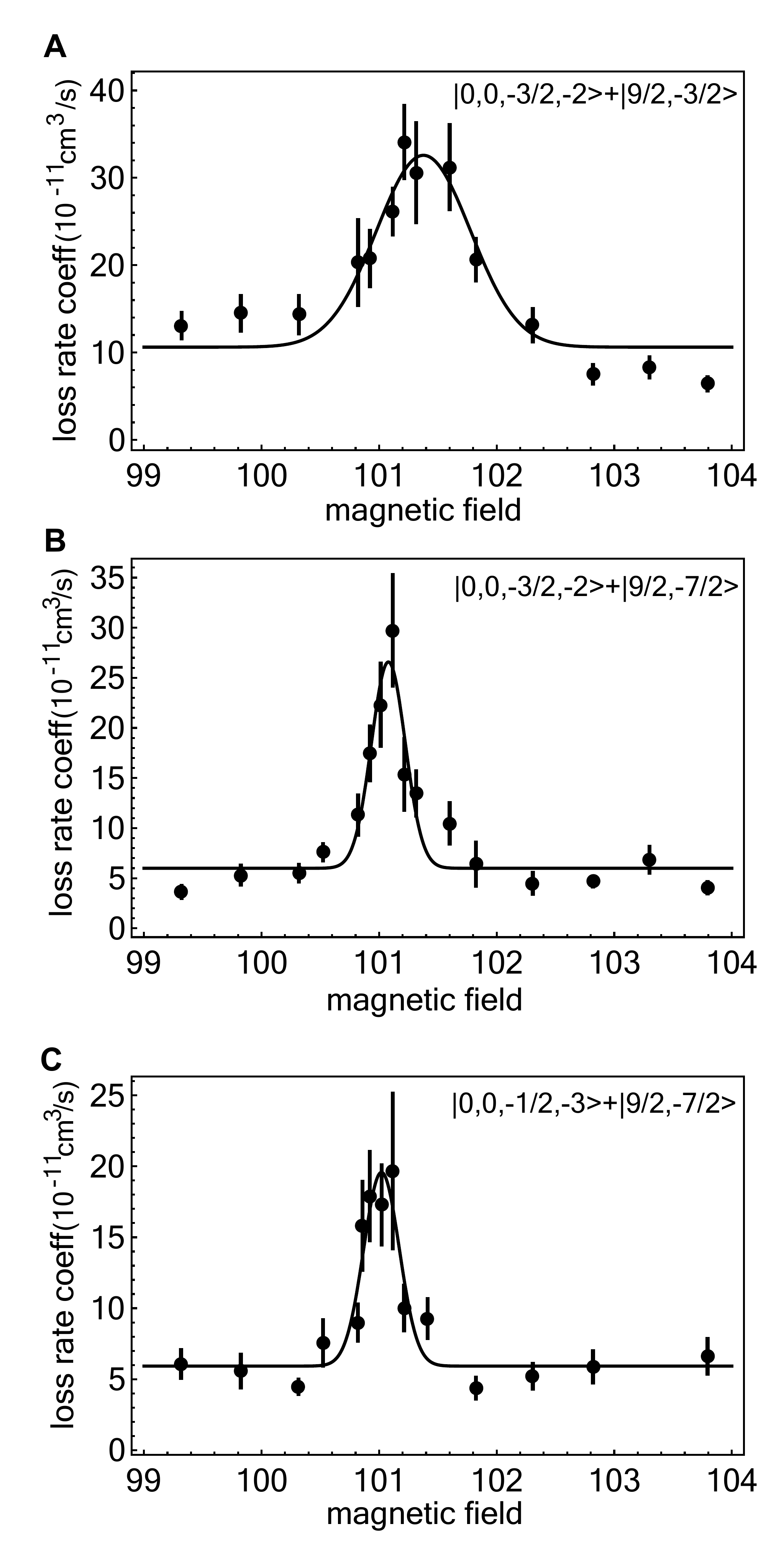}
\caption{\textbf{Observations of the atom-molecule Feshbach resonances in the loss rate coefficients}. The loss rate coefficients are plotted as a function of the magnetic field. The collision channels are (\textbf{A}) $|0,0,-3/2,-2\rangle$+$|9/2,-3/2\rangle$, (\textbf{B}) $|0,0,-3/2,-2\rangle$+$|9/2,-7/2\rangle$, and (\textbf{C}) $|0,0,-1/2,-3\rangle$+$|9/2,-7/2\rangle$. For these three channels, the resonantly enhanced loss rate coefficients at about 101 G provide the clear evidences of the atom-molecule Feshbach resonances. The solid lines are the Gaussian fits. Error bars represent $\pm1$ s.d.}%
\label{fig3}%
\end{figure}

In the experiment, we first coarsely scan the magnetic field with a step of about 1 G. If a signature of Feshbach resonance appears, we reduce the scanning step to about 0.1-0.2 G. We find the loss rate coefficients are different for various collision channels. For each channel, in most cases, the loss rate coefficients do not change significantly in the magnetic field range. However, in the $|0,0,-3/2,-2\rangle+|9/2,-3/2\rangle$, $|0,0,-3/2,-2\rangle+|9/2,-7/2\rangle$, and $|0,0,-1/2,-3\rangle+|9/2,-7/2\rangle$ collision channels , the loss rate coefficients show prominent features at about 101 G, as shown in Fig. 3. We attribute these loss features to the resonant enhancement of the inelastic collisions due to the \emph{s}-wave atom-molecule Feshbach resonance. The resonance positions and widths obtained by the Gaussian fits are listed in Table 1.

\begin{table} [pth]
\centering
\begin{tabular}{c|c|c}
\hline \hline
incoming collision channel & $B_0$ (G) & $\Delta B$ (G)    \\ \hline

 $|0,0,-3/2,-2\rangle+|9/2,-3/2\rangle$ & 101.4 & 0.6        \\ \hline
 $|0,0,-3/2,-2\rangle+|9/2,-7/2\rangle$ & 101.1 & 0.2        \\ \hline
 $|0,0,-1/2,-3\rangle+|9/2,-7/2\rangle$ & 101.0 & 0.2         \\

\hline \hline
\end{tabular}
\caption{The Feshbach resonance position $B_0$ and width $\Delta B$ obtained by the Gaussian fits for the three incoming collision channels. }
\label{Table}
\end{table}

The observations of the Feshbach resonances allow us to compare with the density of resonant states estimated from the statistical model. For the $^{23}$Na$^{40}$K+$^{40}$K collision studied in our experiment, neglecting the nuclear spins, the density of resonant states are estimated to be about 1.22 per mK  \cite{Mayle2012}. If the nuclear spins are considered, the density of resonant states is multiplied by a factor of $N_{\rm{nuc}}$, where $N_{\rm{nuc}}$ is the number of spin states that conserve the total magnetic quantum number. Assume the short-range physics does not change with the magnetic field, the resonance spectrum is probed with a rate of the Zeeman shift of the scattering channel  \cite{Mayle2012}. These arguments predict many \emph{s}-wave resonances with an average spacing of about 1 G for the three collision channels. However, in the approximate 4-G-wide magnetic field range, we observe only a single resonantly enhanced loss feature in each collision channel. This suggests that the density of resonant states may be not as large as the statistical model predicts. One possible reason is including the nuclear spin by simply multiplying $N_{\rm{nuc}}$  is not appropriate. Besides, the magnetic field dependence of the energy of the triatomic bound states is not considered in the model. Note that we cannot exclude the possibility that some resonant states are not observed because of the weak coupling between the bound states and the scattering states.

It is interesting to compare the widths of the observed Feshbach resonances and the number of open channels $N_0$, which is precisely determined for each incoming collision channels. The statistical model in Ref. \cite{Mayle2012} predicts that the width of the resonance is proportional to the number of open channels. For the $|0,0,-3/2,-/2\rangle+|9/2,-3/2\rangle$ incoming channel, we have $N_0=15$. For the $|0,0,-3/2,-2\rangle$ or $|0,0,-1/2,-3\rangle+|9/2,-7/2\rangle$ incoming channels, we have  $N_0=7$ or 6. The fitted width of the first resonance is larger than the widths of the latter two resonances. This agrees with the predictions. However, the three resonances are all resolvable, which is not consistent with the predictions that for $N_0>2\pi$, the resonances start to overlap and may be not resolvable.

In conclusion, we have observed magnetically tunable Feshbach resonances in ultracold collisions between $^{23}$Na$^{40}$K ground-state molecules and $^{40}$K atoms.
The unprecedented controllability offered by the ultracold molecule allows us to observe the atom-molecule resonances in ultracold collisions involving the heavy molecules. Different from the molecular system where resonances are observed in cold collisions involving light particles, in our system, the reduced mass of 24.4 atomic units is pretty large, and the temperature of less than 1 $\mu$K is lower by at least 4 orders of magnitude. In such a heavy and ultracold system, there may be a lot of resonances in a magnetic field range of a few hundred Gauss, albeit the density of resonant states may be not as high as the statistical model estimates. The magnetic field range in the current experiment is close to the atomic Feshbach resonance, and we expect more atom-molecule resonances can be observed in different magnetic fields. The observation of more resonances will enable the study of the quantum chaos in the ultracold molecular collisions \cite{Mayle2012}.

The observed atom-molecule Feshbach resonances at ultracold temperatures probe the short-range resonance spectrum with an unprecedented solution, and provide valuable information of the PES governing the collision dynamics. However, understanding these resonances quantitatively is quite a challenge to the theory. The accuracy of the PES has to be significantly improved. Although $^{23}$Na$^{40}$K+$^{40}$K collision is nonreactive, a full treatment of the collision dynamics must use reactive formalism, because the closed channels asymptotic to the reaction channels are locally open at short range \cite{Croft2017}. Moreover, nuclear spins, hyperfine interactions and external magnetic fields should be considered in the calculation. The observation of Feshbach resonances opens up the exciting possibility of studying resonantly interacting atom-molecule mixture. The resonances may also be employed to create ultracold triatomic molecules from the atom-molecule mixture using the magnetic association \cite{chin2010}.

This work was supported by the National  Key R\&D Program of China (under Grant No. 2018YFA0306502), National Natural Science Foundation of China (under Grant No.11521063), and the
Chinese Academy of Sciences.

\clearpage

\section*{Supplementary materials}

\section*{Creation of weakly bound Feshbach molecules}

Our experiment starts with approximately $3\times10^5$ $^{23}$Na and $1.6\times10^5$ $^{40}$K atoms confined in a crossed-beam optical dipole trap at a temperature of about 500 nK. The trap frequencies for $^{40}$K atoms are $2 \pi\times (247,237,69)$ Hz. The details of the preparation of the ultracold mixture can be found in our previous work \cite{Rui2017,Zhu2017}. The atomic Feshbach resonance between the $|f,m_f\rangle_{\rm{Na}}=|1,1\rangle$ and $|f,m_f\rangle_{\rm{K}}=|9/2,-7/2\rangle$ states at about 110 G \cite{Park2012} is employed to
create weakly bound Feshbach molecules. The Feshbach molecules are created by two-photon Raman photoassociation method.
We use two blue-detuned Raman light fields to couple the free scattering state to the Feshbach molecule state with a single-photon detuning of $\Delta\approx2\pi\times251$
GHz. The Raman Rabi frequency for the $|9/2,-5/2\rangle\rightarrow |9/2,-7/2\rangle$ transition is about $2\pi \times90$ kHz. The efficiency of Raman photo-association is about 15\%. Typically, we create about $2.3\times10^{4}$ Feshbach molecules. After the formation of the Feshbach molecules, the Na atoms are removed by transferring the  $|1,1\rangle$ to $|2,2\rangle$ state and then blasting out by resonant light.

\section*{Stimulated Raman adiabatic passage}

The level scheme of the stimulated Raman adiabatic passage is shown in Fig. S1. The pump laser is a tapered amplified diode laser (805 nm) and the Stokes
laser is a tapered amplified second-harmonic generation laser (567 nm). The
pump laser and the seed light of the Stokes laser are locked to an ultralow
expansion (ULE) cavity with a finesse of about 7500 for 805 nm and 9100 for
1134 nm, respectively. The laser linewidth are estimated to be about 1 kHz.

\begin{figure}[pth]
\centering
\renewcommand\thefigure{S1}
\includegraphics[width=8cm]{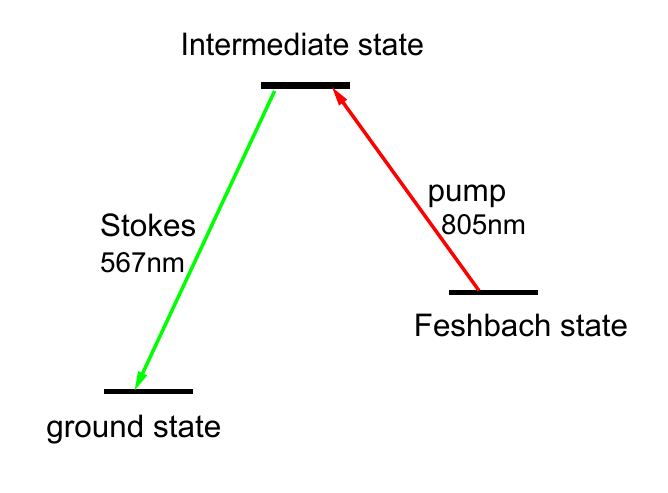}
\caption{The $^{23}$Na$^{40}$K molecules are transferred from the Feshbach state to the rovibrational ground state by the stimulated Raman adiabatic passage. The wavelengths of pump laser and the Stokes laser are 805 nm and 567 nm respectively.}%
\label{fig1}%
\end{figure}

To transfer the Feshbach molecule to the rovibrational ground state, we employ the mixture
of $B^{1}\Pi$ $|v=12,J=1\rangle$ and $c^{3}\Sigma$ $|v=35,J=1\rangle$ molecule
electronic excited states \cite{Park2015} as the intermediate states. The hyperfine structure of the electronic excited state manifold has
been extensively investigated in Ref. \cite{Park2015b,Ishikawa1992}. Starting from the Feshbach molecule, the different hyperfine levels of the ground states can be populated by
choosing the different hyperfine levels of the electronic excited states and the polarizations of the Stokes and pumping lasers. In Table S1, we list the hyperfine ground states used in this work, and the corresponding intermediate states for the STIRAP and the required polarization of the Stokes lasers and pumping lasers. The hyperfine levels of the electronic excited states may be labelled by the quantum numbers $F_1=I_{\rm{Na}}+J$ and $F=F_1+I_{\rm{K}}$ \cite{Ishikawa1992,Park2015b}. Since for the electronic excited state the hyperfine interaction between the nuclear spin of $^{40}$K atom and the electron spin is small, the quantum numbers $F_1$, $m_{F_{1}}$,  and $m_F^{e}=m_{F_{1}}+m^{e}_{I_{\rm{K}}}$ are approximately good quantum numbers.

The $^{23}$Na$^{40}$K molecule in the rovibrational ground has 36 hyperfine states, labelled by $|0,0,m_{I_{\rm{Na}}},m_{I_{\rm{K}}}\rangle$ with $m_{I_{\rm{Na}}}=-3/2,...,3/2$ and $m_{I_{\rm{K}}}=-4,...,4$. At $B=100$ G, the energy difference between $|0,0,m_{I_{\rm{Na}}},m_{I_{\rm{K}}}\rangle$ and $|0,0,m_{I_{\rm{Na}}}-1,m_{I_{\rm{K}}}\rangle$ is approximately 112 kHz, and the energy difference between $|0,0,m_{I_{\rm{Na}}},m_{I_{\rm{K}}}\rangle$ and $|0,0,m_{I_{\rm{Na}}},m_{I_{\rm{K}}}+1\rangle$ is approximately 24 kHz. The hyperfine structures of the ground state molecules are shown in Fig.1 of the main text.

\begin{table*} [pth]
\renewcommand\thetable{S1}
\begin{tabular}{c|c|c|c}
\hline \hline
hyperfine ground state & intermediate state & $m_F^{e}=$&required polarization   \\
                       &                         & $m_{F_{1}}+m^{e}_{I_{K}}$   &Stokes/pump   \\ \hline
 $|0,0,-1/2,-2\rangle$ & $F_{1}\approx1/2,m_{F_{1}}\approx-1/2,m^{e}_{I_{K}}\approx-2$ & -5/2 &$\pi/\pi$       \\ \hline
 $|0,0,-1/2,-3\rangle$ & $F_{1}\approx1/2,m_{F_{1}}\approx-1/2,m^{e}_{I_{K}}\approx-3$ & -7/2&$\pi/\sigma^{+}$  \\    \hline
 $|0,0,-3/2,-1\rangle$ & $F_{1}\approx3/2,m_{F_{1}}\approx-3/2,m^{e}_{I_{K}}\approx-1$ & -5/2&$\pi/\pi$         \\  \hline
 $|0,0,-3/2,-2\rangle$ & $F_{1}\approx3/2,m_{F_{1}}\approx-3/2,m^{e}_{I_{K}}\approx-2$ & -7/2&$\pi/\sigma^{+}$ \\
\hline \hline
\end{tabular}
\caption{The hyperfine ground states, the corresponding intermediate states, and the required polarizations of the lasers.}
\label{Table}
\end{table*}

In our experiment, the Stokes laser is $\pi$ polarized. The required polarization of the pump laser
is $\pi$ for the $m^{e}_F=-5/2$ intermediate state or $\sigma^{+}$ for $m^{e}_F=-7/2$ intermediate state. Due to the limitation of the experimental configuration, both Stokes
and pump lasers propagate perpendicular to the magnetic field. In this configuration, the $\pi$
polarization is achieved by setting the polarization of the laser parallel to the magnetic field. The $\sigma^{+}$ polarization is
applied by setting the polarization of the laser to $\hat{x}=(\sigma^{+}+\sigma^{-})/\sqrt{2}$ linear polarization. The $\sigma^{-}$ component will also couple the Feshbach molecule to the $m_F^{e}=-3/2$ hyperfine level of the electronic excited state. However, a detailed calculation shows the Rabi frequency between the Feshbach state and the $m_F^{e}=-7/2$ intermediate state is about 4 times larger than the Rabi frequency between the Feshbach state and the $m_F^{e}=-3/2$ hyperfine state. Therefore, the STIRAP with a $\pi/\hat{x}$ polarization will dominantly transfer the Feshbach molecule to the desired hyperfine ground state via the $m_F^{e}=-7/2$ intermediate state. The Rabi frequencies for Stokes and pump light are both about 1 MHz and the efficiency of round-trip STIRAP is about 15-20\%. We use a low Rabi frequency to ensure a high degree purity of the hyperfine ground states.

\section*{Loss rate coefficients}

\begin{figure*}[pth]
\centering
\renewcommand\thefigure{S2}
\includegraphics[width=16cm]{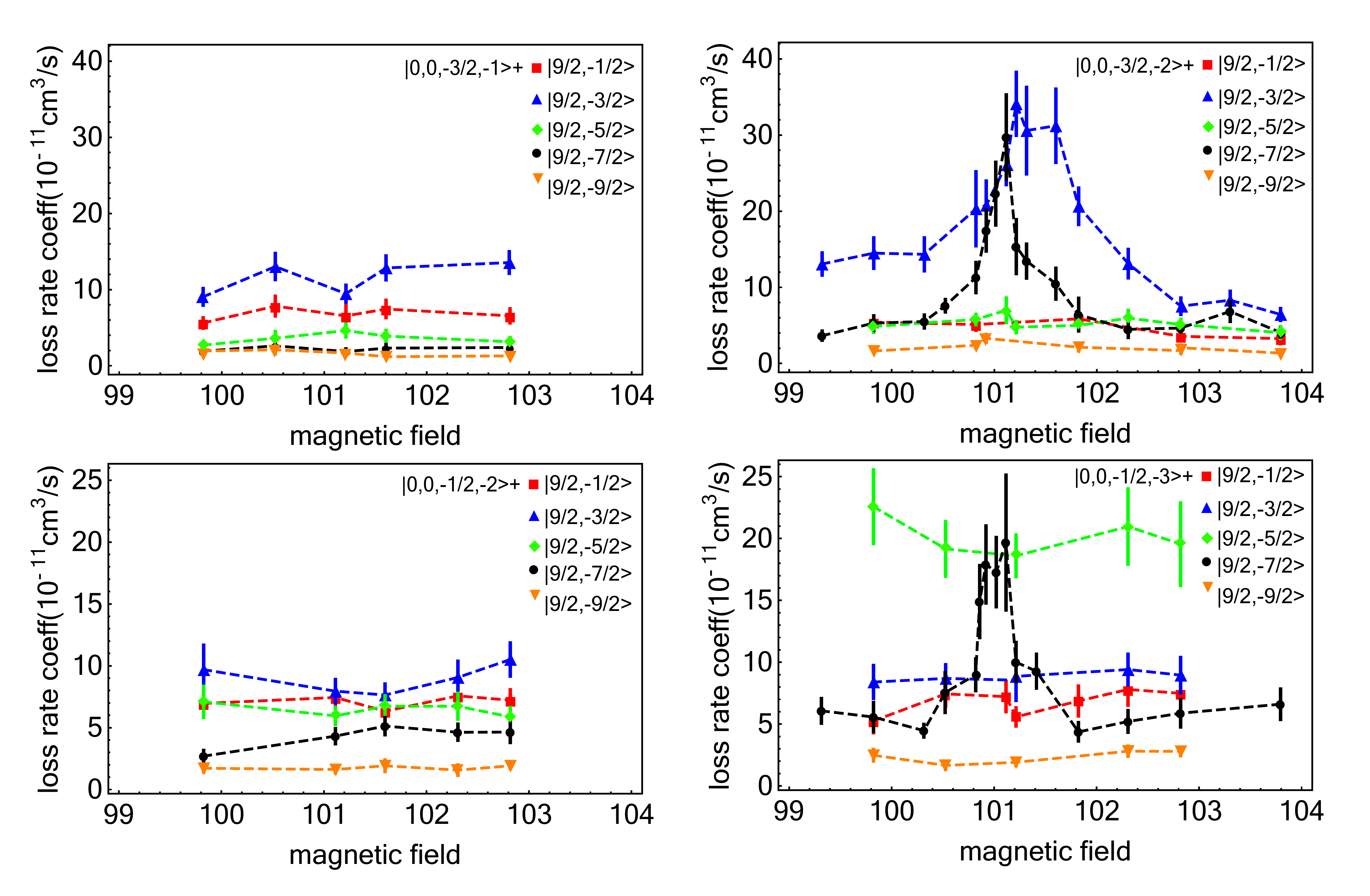}
\caption{The measured loss rate coefficients as a function of the magnetic field in all the twenty collision channels. Error bars represent $\pm1$ s.d.}%
\label{fig2}%
\end{figure*}

The loss rate coefficients as a function the magnetic field in all the twenty collision channels are shown in the Fig. S2. Besides the three narrow resonances, the loss rate coefficients in  the $|0,0,-1/2,-3\rangle+|9/2,-5/2\rangle$ collision channel are unusually large in the whole magnetic field range. Further experimental work is needed to identify whether this is a broad Feshbach resonance.


\begin{thebibliography}{40}

\bibitem{Herschbach2009}
D.~Herschbach, {\it Faraday Discussions\/} {\bf 142}, 9 (2009).

\bibitem{chin2010}
C.~Chin, R.~Grimm, P.~Julienne, E.~Tiesinga, {\it Rev. Mod. Phys.\/} {\bf 82},
  1225 (2010).

\bibitem{Skodje2000}
R.~T. Skodje, {\it et~al.\/}, {\it Phys. Rev. Lett.\/} {\bf 85}, 1206 (2000).

\bibitem{Qiu2006}
M.~Qiu, {\it et~al.\/}, {\it Science\/} {\bf 311}, 1440 (2006).

\bibitem{Henson2012}
A.~B. Henson, S.~Gersten, Y.~Shagam, J.~Narevicius, E.~Narevicius, {\it
  Science\/} {\bf 338}, 234 (2012).

\bibitem{Wang2013}
T.~Wang, {\it et~al.\/}, {\it Science\/} {\bf 342}, 1499 (2013).

\bibitem{Kim2015}
J.~B. Kim, {\it et~al.\/}, {\it Science\/} {\bf 349}, 510 (2015).

\bibitem{Yang2015}
T.~Yang, {\it et~al.\/}, {\it Science\/} {\bf 347}, 60 (2015).

\bibitem{Vogels2015}
S.~N. Vogels, {\it et~al.\/}, {\it Science\/} {\bf 350}, 787 (2015).

\bibitem{Bergeat2015}
A.~Bergeat, J.~Onvlee, C.~Naulin, A.~van~der Avoird, M.~Costes, {\it Nature
  Chemistry\/} {\bf 7}, 349 (2015).

\bibitem{Klein2017}
A.~Klein, {\it et~al.\/}, {\it Nature Chemistry\/} {\bf 13}, 35 (2017).

\bibitem{Chefdeville2013}
S.~Chefdeville, {\it et~al.\/}, {\it Science\/} {\bf 341}, 1094 (2013).

\bibitem{Perreault2017}
W.~E. Perreault, N.~Mukherjee, R.~N. Zare, {\it Science\/} {\bf 358}, 356
  (2017).

\bibitem{Carr2009}
L.~D. Carr, D.~DeMille, R.~V. Krems, J.~Ye, {\it New J. Phys.\/} {\bf 11}, 1367
  (2009).

\bibitem{Quemener2012}
G.~Qu\'{e}m\'{e}ner, P.~S. Julienne, {\it Chem. Rev.\/} {\bf 112}, 4949 (2012).

\bibitem{Bohn2002}
J.~L. Bohn, A.~V. Avdeenkov, M.~P. Deskevich, {\it Phys. Rev. Lett.\/} {\bf
  89}, 203202 (2002).

\bibitem{Mayle2012}
M.~Mayle, B.~P. Ruzic, J.~L. Bohn, {\it Phys. Rev. A\/} {\bf 85}, 062712
  (2012).

\bibitem{Tscherbul2006}
T.~V. Tscherbul, R.~V. Krems, {\it Phys. Rev. Lett.\/} {\bf 97}, 083201 (2006).

\bibitem{Hummon2011}
M.~T. Hummon, {\it et~al.\/}, {\it Phys. Rev. Lett.\/} {\bf 106}, 053201
  (2011).

\bibitem{Frye2016}
M.~D. Frye, M.~Morita, C.~L. Vaillant, D.~G. Green, J.~M. Hutson, {\it Phys.
  Rev. A\/} {\bf 93}, 052713 (2016).

\bibitem{Tscherbul2009}
T.~V. Tscherbul, Y.~V. Suleimanov, V.~A. andR V~Krems, {\it New J. Phys\/} {\bf
  11}, 055021 (2009).

\bibitem{Ni2008}
K.-K. Ni, {\it et~al.\/}, {\it Science\/} {\bf 322}, 231 (2008).

\bibitem{Takekoshi2014}
T.~Takekoshi, {\it et~al.\/}, {\it Phys. Rev. Lett.\/} {\bf 113}, 205301
  (2014).

\bibitem{Molony2014}
P.~K. Molony, {\it et~al.\/}, {\it Phys. Rev. Lett.\/} {\bf 113}, 255301
  (2014).

\bibitem{Park2015}
J.~W. Park, S.~A. Will, M.~W. Zwierlein, {\it Phys. Rev. Lett.\/} {\bf 114},
  205302 (2015).

\bibitem{Guo2016}
M.~Guo, {\it et~al.\/}, {\it Phys. Rev. Lett.\/} {\bf 116}, 205303 (2016).

\bibitem{Rvachov2017}
T.~M. Rvachov, {\it et~al.\/}, {\it Phys. Rev. Lett.\/} {\bf 119}, 143001
  (2017).

\bibitem{seesselberg2018}
F.~See\ss{}elberg, {\it et~al.\/}, {\it Phys. Rev. A\/} {\bf 97}, 013405
  (2018).

\bibitem{Ospelkaus2010a}
S.~Ospelkaus, {\it et~al.\/}, {\it Science\/} {\bf 327}, 853 (2010).

\bibitem{Croft2017}
J.~F.~E. Croft, N.~Balakrishnan, B.~K. Kendrick, {\it Phys. Rev. A\/} {\bf 96},
  062707 (2017).

\bibitem{Ziuchowski2010}
P.~S. \ifmmode~\dot{Z}\else \.{Z}\fi{}uchowski, J.~M. Hutson, {\it Phys. Rev.
  A\/} {\bf 81}, 060703 (2010).

\end{thebibliography}

\begin{thebibliography}{10}

\bibitem{Rui2017}
J.~Rui, {\it et~al.\/}, {\it Nature Physics\/} {\bf 13}, 699 (2017).

\bibitem{Zhu2017}
M.-J. Zhu, {\it et~al.\/}, {\it Phys. Rev. A\/} {\bf 96}, 062705 (2017).

\bibitem{Park2012}
J.~W. Park, {\it et~al.\/}, {\it Phys. Rev. A\/} {\bf 85}, 051602 (2012).

\bibitem{Park2015}
J.~W. Park, S.~A. Will, M.~W. Zwierlein, {\it Phys. Rev. Lett.\/} {\bf 114},
  205302 (2015).

\bibitem{Park2015b}
J.~W. Park, S.~A. Will, M.~W. Zwierlein, {\it New J. Phys.\/} {\bf 17}, 075016
  (2015).

\bibitem{Ishikawa1992}
K.~Ishikawa, T.~Kumauchi, M.~Baba, , H.~Kat\^{o}, {\it Journal of Chem. Phys.\/}
  {\bf 96}, 6423 (1992).

\end{thebibliography}
\end{document}